\documentclass[twocolumn,showpacs,preprintnumbers,amsmath,amssymb,prb]{revtex4-1}
\usepackage{graphicx}
\usepackage{dcolumn}
\usepackage{bm}
\usepackage{epsfig}
\usepackage{subfigure}
\usepackage{graphicx}
\usepackage{mathcomp}
\usepackage{sidecap}
\usepackage{url}

\newcommand{\be}{\begin{equation}}
\newcommand{\ee}{\end{equation}}

\newcommand{\bea}{\begin{eqnarray}}
\newcommand{\eea}{\end{eqnarray}}

\begin{document}

\title{ Multiple liquid-liquid critical points and density anomaly in core-softened potentials}
\author{Marco Aur\'elio A. Barbosa}
\email{aureliobarbosa@unb.br}
\affiliation{Programa de P\'os-Graduac\~ao em Ci\^encia de Materiais, Faculdade UnB Planaltina, Universidade de Bras\'ilia, Planaltina-DF, Brazil}
\author{Evy Salcedo}
\email{esalcedo@fsc.ufsc.br}
\affiliation{Departamento de F\'isica, Universidade Federal do Santa Catarina, Florian\'opolis-SC, Brazil}
\author{Marcia C. Barbosa}
\email{marcia.barbosa@ufrgs.br}
\affiliation{Instituto de F\'isica, Universidade Federal do Rio Grande do Sul, Porto Alegre-RS, Brazil}

\begin{abstract}
The relation between liquid-liquid phase transitions and waterlike density anomalies in core-softened potentials of fluids was investigated in an exactly solvable one dimensional lattice model and a in a three dimensional fluid with fermi-like potential, the latter by molecular dynamics. Both systems were shown to present three liquid phases, two liquid-liquid phase transitions closely connected to two distinct regions of anomalous density increase. We propose that an oscillatory behavior observed on the thermal expansion coefficient as a function of pressure can be used as a signature of the connection between liquid-liquid phase and density.
\end{abstract}
\maketitle

The phase behavior of single component systems as particles interacting via the o-called core-softened (CS) potentials has received attention since
the pioneering work of Stell and Hemmer proposing the possibility of a second critical point in addition to the usual liquid-gas critical point~\cite{He70}. These potentials exhibit a repulsive core with a softening region with a shoulder or a ramp~\cite{He70,St72,Ja98} which are analytically and computationally tractable while being capable of retaining the qualitative features of the real fluid systems. 
Furthermore, Debenedetti and collaborators~\cite{De91} using thermodynamic arguments showed that the 
isobaric thermal expansion coefficient ($\alpha$) of these potentials might have
an anomalous negative value and consequently a region where density \textit{increases} with temperature. Since for high temperatures density \textit{decreases} with temperature these potentials can exhibit a temperature of maximum density (TMD) connecting the two regions.
 
The thermodynamic anomalies predicted by these models occur in liquids such as Te~\cite{Th76}, 
Ga, Bi~\cite{Handbook}, S~\cite{Sa67,Ke83}, liquid water~\cite{Ke75}, and
Ge$_{15}$Te$_{85}$~\cite{Ts91}, and were found in simulations for
silica~\cite{An00,Po97}, silicon~\cite{Sa03}, and
BeF$_2$~\cite{An00}. In addition experiments in phosphorous indicate the presence of a liquid-liquid phase transition~\cite{Ka00} and similar transitions were observed by simulations in models of silica~\cite{Vo01b}, silicon~\cite{Sa03} and liquid water~\cite{Po92}.

In the particular case of water the hypothesis of the existence of two
liquid phases has been indirectly supported by experimental results in confined systems~\cite{Xu05}.
In spite of the limit of 235 K below which water cannot be found in the liquid phase without
crystallization, two amorphous phases, 
a low density amorphous phase and a high density amorphous phase, were observed at much lower temperatures and
their relation to a metastable liquid-liquid phase transition was argued~\cite{Mi02}.
More recently a third amorphous phase, the very high density amorphous phase, has been observed~\cite{Bo06} what suggests the possibility of the existence of  the very high density liquid phase.

Therefore the issue of a liquid-liquid phase transition  and its connection with the existence of a region in the pressure-temperature phase diagram where the density decreases with the decrease of temperature is itself an interesting topic. It is accepted that the presence of two accessible length scales in the potential allow for the system to have two liquid phases and a density anomaly. The accessibility is the ingredient that explains why a density anomaly derived in 1D does not necessarily hold at higher dimensions and why a density anomaly derived for a smooth potential might be lost if the slope linking the two length scales becomes infinite~\cite{Fr01,Ol08b,Fo08}.

In this letter we show, by means of an exactly solvable 1D model and numerical simulations of a similar 3D potential, that a three length scales potential might exhibit three critical points and two density anomalous regions if the different length scales would be accessible. To the best of our knowledge this is the first time that core softened potentials are shown to have three critical points and two temperature of maximum density lines. 
In addition we propose propose a new way to identify if a liquid-liquid critical point would be present by exploring the behavior of $\alpha$ for temperatures above the critical temperature. We show that, for systems in which there is a density anomalous region close to criticality, $\alpha$ exhibits a peculiar behavior diverging to $+\infty$ for pressures above the critical pressure and to  $-\infty$ for pressures below the critical pressure. 

\begin{figure}[ht]
\begin{center}
\includegraphics[clip,width=7.5cm]{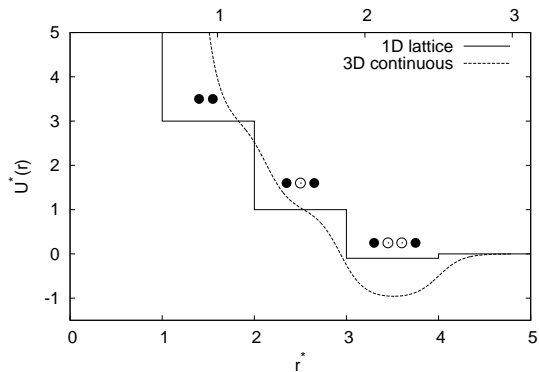}
\caption{Representation of the pair potentials investigated on this paper: a one dimensional lattice fluid and an isotropic three dimensional continuous core softened model with three scales of interaction. Top (bottom) horizontal axis used to represent continuous (lattice) system. } 
\label{fig:potential-repulsive}
\end{center}
\end{figure}

The core-softened potentials we analyze are illustrated in Fig.~\ref{fig:potential-repulsive} in units of length $\sigma$ and of energy $\epsilon$. 
They are composed by a hard core of diameter $\sigma$, two repulsive shoulders and an attractive well with depth $-\lambda\epsilon$ where we employ $\lambda=1/10$ for the 1D case and $\lambda=1$ for the 3D case. Both potential are detailed below.

We first investigate the 1D to obtain analytical insights and to become familiar with the properties that will also appear in the 3D case. 
The linear lattice has length $L$ and $N$ sites separated by distance $\sigma=L/N$.
The exact form of the Gibbs free energy derived in the framework of the Takahashi method~\cite{Ta66} is given by
\be
G(T,P,N) = -N k_B T \ln \left [ \sum_{x=1}^{\infty} e^{-\beta h(r;P) } \right ],
\label{eq:G}
\ee
where $\beta=1/(k_B T)$ with $k_B$ the Boltzmann constant, and $h(r;P) = U(r) + Pr$ the microscopic
pair enthalpy, with $U(r)$ being the interaction energy between neighbor molecules. Expression~(\ref{eq:G}) is calculated using $U(\sigma)=3\epsilon$, $U(2\sigma)=\epsilon$,  $U(3\sigma)=-\epsilon/10$, and $U(r)=0$ for $r>3\sigma$. 

The minimization of the Eq.~(\ref{eq:G}) in the ground state results in three configurations depicted in Fig.~\ref{fig:potential-repulsive}: a low density liquid (LDL) phase, a high density liquid (HDL) phase and a very high density liquid (VHDL) phase, besides a gas phase (G), not illustrated. Coexistence between these phases allows for three ground state phase transitions (GSTP) to occur at reduced pressures ($P^*=P\sigma /\epsilon$) $P^*_{G-LDL} = 0 $; $P^*_{LDL-HDL}=1.1$, and $P^*_{HDL-VHDL}= 2.0$. In what follows temperature is reduced as $T^* = k_B T / \epsilon$.

\begin{figure}[ht]
\begin{center}
\includegraphics[clip,width=7.5cm]{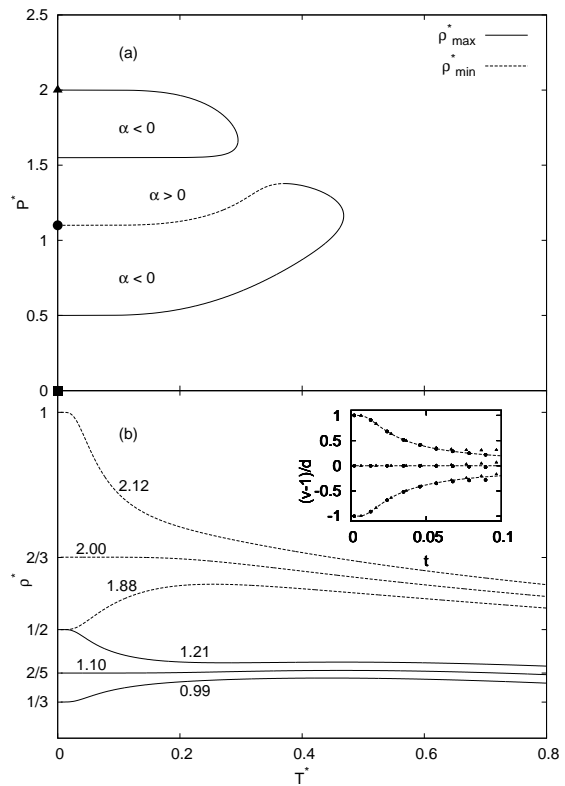}
\caption{ (a) Pressure vs. temperature phase diagram for the 1D case. Continuous and dotted lines indicate the TmD and the TMD lines, while symbols locate the ground state phase transitions. (b) Density as a function of temperature, for various pressures. The inset shows the reduced volume as a function of the reduced temperature near the two critical points, according to~(\ref{eq:vs-v}). }
\label{fig:density} 
\end{center}
\end{figure} 

\begin{figure}[ht!]
\begin{center}
\includegraphics[clip,width=7.5cm]{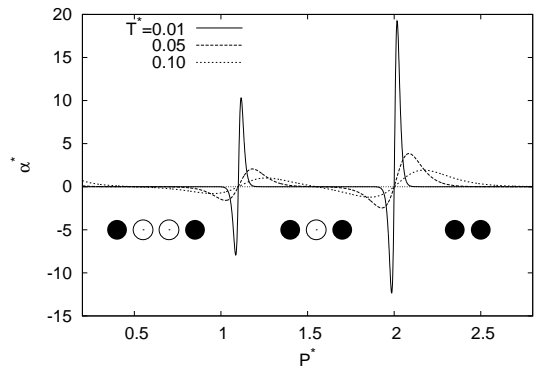}
\caption{Thermal expansion coefficient ($\alpha^*=\epsilon \alpha$) as a function of pressure at fixed temperatures, for the 1D case.} 
\label{fig:alpha}
\end{center}
\end{figure}

The full phase diagram for the 1D case is illustrated on Fig.~\ref{fig:density} (a) with symbols indicating the three GSPT described in the last paragraph. From the LDL-HDL critical point (circle) emerges a line of temperature of minimum density (TmD) followed by a line of temperature of maximum density (TMD),
while from the HDL-VHDL critical (triangle) point appears a TMD. These lines separate the region where the thermal expansion coefficient  is positive from the regions where $\alpha$ is negative. Fig.~\ref{fig:density} (b) illustrates the  density as function of the temperature for various pressures, some of them crossing the TMD and the TmD lines. 

Further comprehension on the relation between density anomaly and GSPT can be gained by visualizing $\alpha$ as a function of pressure (at fixed temperature), as shown in Fig.~\ref{fig:alpha}. In usual systems $\alpha \rightarrow \infty$ as the critical point is approached from any path in the pressure-temperature plane.  The peculiarity of models in which the criticality is associated with the density anomaly is that below the critical pressure $P_c$ the sign of $\alpha$ is negative, while for $P>P_c$ its sign is positive. Consequently $\alpha$ displays an oscillatory behavior as a function of pressure, with $\alpha(P \rightarrow P_c^-) \rightarrow - \infty$ and $\alpha(P \rightarrow P_c^+) \rightarrow + \infty$, which is a signature of the connection between the TMD line (or TmD) and the critical point~\cite{barbosa11jcp}.

The relation between criticality and anomalous density behavior can be rationalized by looking at the thermodynamic close to the GSPT within the approximation adopted in Ref.~\cite{fiore11:prl}. For a lattice model with core softened interaction the reduced Gibbs free energy per particle in the vicinity of any GSPT can be written in terms $r_1$ and $r_2$, the average distances between the particles in the two coexisting phases, leading to 
\be
 g(t,p) = h_c + p - t \ln \left [  2 \cosh \left ( \frac{pd}{t} \right )  \right ],
\label{eq:g}
\ee
where $g =  G /(N P_c r_c)$, $t = k_B T / P_c r_c$, $p = (P - P_c)/P_c $, and $h_c = h(r_1)/(P_c r_c)$. For the LDL-HDL transition
discussed above: $r_1=2\sigma$, $r_2=3\sigma$, $d = (r_1+r_2)/(r_2-r_1)=5$, and $r_c = (r_1+r_2)/2=5\sigma/2$. It follows from~(\ref{eq:g}) that the reduced volume per particle, $v=V/(N r_c)$, can be written as 
\be
v(t,p) = 1 - d \tanh \left (  \frac{pd}{t}  \right ). \label{eq:vs-v}
\ee

 In order to understand the relation between water-like anomalies and phase transitions we investigate the behavior of molecular volume while approaching the GSPT $t=p=0$. Close to this point the value of the reduced volume depends on the path along which criticality is approached, namely
\begin{subequations}
\bea
\lim_{p \rightarrow 0 }  \lim_{t \rightarrow 0} v & = &  \left \{
\begin{array}{lc}
1 - d, & \qquad  p \rightarrow 0^+ \\
1 + d, & \qquad  p \rightarrow 0^- \label{eq:v-lowt-1}
\end{array}
 \right . \\
\lim_{t \rightarrow 0 }  \lim_{p \rightarrow 0} v & = & 1. \label{eq:v-lowt-2}
\eea 
\end{subequations}
The behavior of the reduced volume indicates that slightly below and above the LDL-HDL critical pressure the system is arranged according the LDL and the HDL, respectively, while exactly at the critical pressure the system becomes a mixture of the two phases as the temperature is reduced. Thus, the entropy gained by mixing with the HDL phase explains the density increase observed on the LDL side.

\begin{figure}[ht]
\begin{center}
\includegraphics[clip,width=7.5cm]{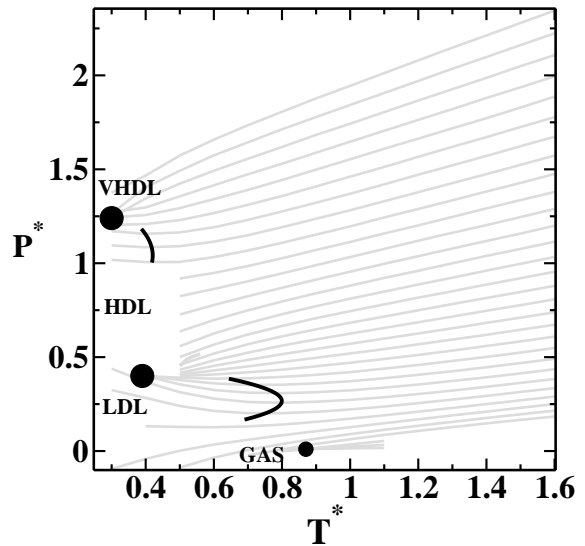}
\caption{Pressure versus temperature phase diagram for the 3D case. The lighter lines are isochores for $\rho^*= 0.12-0.40$, the solid lines the TMD lines and filled circles are the (estimated) location of the critical points.} 
\label{fig:p.vs.t}
\end{center}
\end{figure}

\begin{table}[ht]
\begin{tabular}{|c|c|c|c|}
\hline
\;\;$i\;\;$ & $\;\;\lambda_{i}/\epsilon\;\;$ & $\;\;\upsilon_{i}/\epsilon\;\;$ & $\;\sigma_{i}/\sigma\;$
\tabularnewline \hline
$1$ & $1$ & $0$ & $1$ \tabularnewline \hline
$2$ & $2$ & $1$ & $4/3$ \tabularnewline \hline
$3$ & $2$ & $1$ & $9/5$ \tabularnewline \hline
$4$ & $-1$ & $1$ & $37/15$ \tabularnewline \hline
\end{tabular}
\caption{Coefficients of the core softened pair potential of Eq.~(\ref{eq:potential}). Distances in units of $\sigma=\sigma_1$.}
\label{table1}
\end{table}

Next, we consider if the presence of three critical points and two regions of density anomaly observed for $d=1$ holds for $d>1$. To this end, we perform molecular dynamic  simulations, using the HOOMD-blue package~\cite{And08,hoomd_URL}, at constant temperature (NVT) and constant pressure (NPT) in a three dimensional  system composed of $2048$ particles in a cubic box with periodic boundary conditions,  interacting through a continuous pair potential obtained by the addition of $3$ different Fermi-Dirac distributions,
\begin{equation}
U=\epsilon
\sum_{i=1}^{4}\frac{\lambda_{i}}{\upsilon_{i}+\exp\left(\frac{r-\sigma_{i}}{\omega}\right)}
\label{eq:potential}
\end{equation}
with $\omega = 0.08\sigma$ and the other coefficients given in Table~\ref{table1}. The potential  illustrated as the smooth curve in
Fig.~\ref{fig:potential-repulsive} was chosen to mimic the three length scales and energies scales employed in the one dimensional case. The three length scales are identified by the change of the slope in the force~\cite{De91}. The accessibility of the three length scales and the presence of the density anomalous region in the stable region of the pressure-temperature phase diagram is obtained by making the forces smooth and positive~\cite{Ba11}. Units are reduced as in the 1D case, except for pressure and particle density which are reduced as $P^*=P \sigma^3/\epsilon$ and $\rho^*=\sigma^3 \rho$.

\begin{figure}[ht]
\begin{center}
\includegraphics[clip,width=7.5cm]{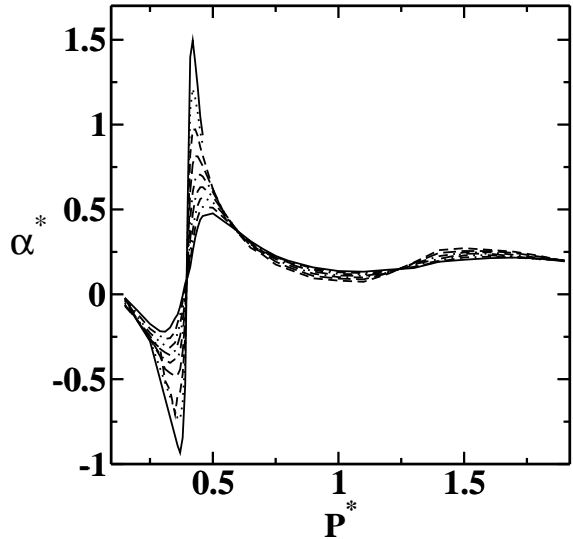}
\caption{Thermal expansion coefficient versus pressure for
$T^*=0.48-0.68$.} 
\label{fig:alpha.vs.p}
\end{center}
\end{figure}

The pressure versus temperature phase diagram is shown in Fig.~\ref{fig:p.vs.t}. At low temperatures, the system presents gas, LDL, HDL, and VHDL phases. Gray lines are the isochores $\rho^*= 0.12-0.40$ and filled circles indicate gas-LDL, LDL-HDL, and HDL-VHDL critical points. Similarly to what happens in the 1D case, in the 
vicinity of both LDL-HDL and HDL-VHDL critical points there are TMD lines located in the stable region of the pressure-temperature phase diagram. 
In both cases the TMD lines seems to approach the critical pressures~\cite{Sa11}, allowing us to test if the oscillatory behavior
observed in $\alpha$ in the one dimensional case can also occur in the 3D case.
This behavior is observed in Fig.~\ref{fig:alpha.vs.p}, where the thermal expansion coefficient is drawn as a function of the pressure for different fixed temperatures $T^*=0.48-0.68$ in the vicinity of the LDL-HDL critical point. As in the 1D case $\alpha$ goes to $-\infty$ as $P\rightarrow P_c^-$
while it diverges to  $+\infty$ as $P\rightarrow P_c^+$ at the critical temperature.

In this paper we used an exactly solvable one dimensional lattice model and molecular dynamic simulations of a three dimensional fluid model to show that three length scales core-softened potential can be designed to exhibit three liquid phases (LDL, HDL and VHDL), two liquid-liquid critical points and two density anomalous regions. In addition, we propose that the oscillatory behavior observed in the thermal expansion coefficient close to criticality can be considered as a signature of the connection between a liquid-liquid critical point and a region of anomalous density increase, in the studied models. In the particular case of water the hypothesis of a liquid-liquid critical point is still under debate due to the impossibility to experimentally probe the system in its expected location, at temperatures below the homogeneous nucleation temperature.  We also propose that measurements of $\alpha$ as a function of pressure could be performed at temperatures well above criticality, where the system can be accessed (even though metastable), and used as a tool to test for the presence of a metastable second critical point.

This work is partially supported by CNPq, Capes, INCT-FCx and Universidade Federal de Santa Catarina.
We would like to thank C. E. Fiore and J. Nunes da Silva for usefull discussions.

\bibliographystyle{apsrev}

\end{document}